\documentclass[12pt,a4paper]{article}
\usepackage{graphicx}
\usepackage{epsfig}
\usepackage{epstopdf}
\usepackage{amsmath}
\usepackage{amssymb}
\usepackage{subfig}
\usepackage{ifthen}
\usepackage{alltt}
\usepackage{fancyhdr}
\usepackage{verbatim}
\usepackage{moreverb}
\usepackage{cite}
\usepackage{rotating}
\usepackage{color}
\usepackage{colortbl}
\usepackage{epsfig,amssymb,cite}
\usepackage{multirow}
\usepackage{bigstrut}
\usepackage{epsfig,wrapfig}

\def\e{\begin{equation}}
\def\f{\end{equation}}
\def\_#1{{\bf #1}}

\def\.{\cdot}

\def\l#1{\label{eq:#1}}
\def\r#1{(\ref{eq:#1})}

\begin{document}

\title{Maximizing absorption and scattering by dipole particles}

\def\affil#1{\begin{itemize} \item[] #1 \end{itemize}}

\author{Sergei Tretyakov}



\date{\today}

\maketitle

\affil{Department of Radio Science and Engineering, Aalto
University, Finland}

\begin{abstract}

This is a review and tutorial paper which discusses the fundamental
limitations on the maximal power which can be received, absorbed,
and scattered by an electrically small electrically polarizable
particle and infinite periodical arrays of such particles.

\end{abstract}

\medskip

\section{Introduction}

Absorption and scattering of electromagnetic waves by electrically
(optically) small particles are basic, but very important phenomena,
which we encounter in many branches of physics and engineering
(light propagation in atmosphere, small antennas and sensors,
plasmonics, etc.) Especially, \emph{resonant} particles are
interesting and important for applications due to strong
interactions with electromagnetic fields. Most commonly, these are
plasmonic nanoparticles (optical frequencies) and various small
antennas (from radio to terahertz frequencies). On one hand,
understanding of electromagnetic response of small particles appears
to be very simple: After all, we deal with radiation from the
simplest source, a point electric dipole. And indeed, basic
properties of dipole radiators and antennas are well known for very
long time, since the beginning of the 20ies century, and documented
in numerous text books, e.g. \cite{Balanis,Bohren}. In particular,
it is a classical result that the effective area (absorption cross
section) of a small resonant wire antenna is of the order of
$\lambda^2$, where $\lambda$ is the wavelength of the incident
radiation (e.g., \cite{Balanis}). If this effective area would be as
small as the geometrical cross section of the thin metal wire
forming the antenna, no radios would work at frequencies below
microwaves. But on the other hand, it appears that in the optical
literature the fact that a small resonant particle can collect and
absorb much more power than is incident on its geometrical cross
section was recognized only in 1983 \cite{Bohren83,ufn1983}.
Interestingly, the same property has been recognized in 1950ies
mainly in the works of S.M. Rytov in the context of studies of
thermal radiation from finite-size bodies (thermal radiation from a
resonant particle exceeds the black-body limit in the vicinity of
resonances). Recently, with the developments of plasmonics and
metamaterials, basic phenomena  of scattering and absorption in
small resonant particles are again in the focus of research
interests. For example, in paper \cite{perfect2008} it was shown
that a focused laser beam (the focal area of the order of
$\lambda^2$) can be nearly fully reflected by a small resonant
dipole particle, which follows from the fact (see above)  that its
scattering cross section is of the same order.

Likewise, the theory of periodical arrays of small electric dipoles
at normal incidence is a very simple special case of the
well-developed theory of antenna arrays (e.g., \cite{Balanis}). In
particular, if a single small but resonant particle can absorb or
reflect power ``collected'' from the surface of the order
$\lambda^2$, it appears clear that a regular array of such particles
can fully reflect plane waves if the period is smaller than
$\lambda$ (the theory is given e.g. in \cite{modeboo}). This is the
enabling phenomenon in many applications of frequency selective
surfaces (e.g., \cite{Munk}). The Babinet principle tells that an
ideally conducting sheet with periodically arranged electrically
small but resonant holes can be fully transparent at the resonant
frequency (in recent literature, this phenomenon is called
\emph{extraordinary transmission}). In the optical literature,
research on such simple resonant arrays is active even now: For
example, in \cite{perfect_reflector_2013}, this effect of nearly
full reflection from an array of resonant low-loss particles is
demonstrated for an array of dielectric spheres. However, general
considerations on the maximum reflected and received power in such
arrays appeared in the literature only recently
\cite{Pozar_array,Pozar2009}. Paper \cite{Andersen} considers
limiting receiving and scattering performance also of large and
directive antennas.

In most recent literature, there are numerous publications on exotic
and special properties of resonant particles and their arrays,
especially in plasmonics, in connection with the developments of
metasurfaces and metamaterials and new devices (cloaked sensors,
light-trapping structures, etc.) This calls for good understanding
of electromagnetic response of small but resonant structures and
especially on fundamental limitations on their absorptive and
scattering performance. This paper is a tutorial overview with the
focus on fundamental limitations on absorption, extinction, and
scattering cross sections and on most general results which hold for
all small scatterers and planar arrays of such particles. Most
(probably, even all) results and conclusions of this review are
known, but this knowledge is scattered in many publications in
journals and books in various disciplines. There is no attempt to
make a historical overview of the subject, and the reference list is
limited to papers which are necessary to understand this review.
Limitations on electromagnetic response of small particles are
considered also in recent papers \cite{Inigo,Inigo2,alu}. Paper
\cite{younes} studies extreme absorption properties of small
bi-anisotropic particles.

\section{Extracted, scattered, and absorbed powers}

Let us consider a small electrically polarizable particle in free
space, excited by external electric field, which can be assumed to
be approximately uniform over the particle volume. The complex
amplitude of the incident electric field at the position of the
particle we denote as $E_{\rm inc}$.  The linear particle response
we model by its polarizability $\alpha$ defined through \e
p=\alpha\, E_{\rm inc}\f where $p$ is the complex amplitude of the
induced electric dipole moment. The particle can be anisotropic, so
that the induced dipole moment is not necessarily parallel to the
exciting field, but we will need only the ratio between the
amplitudes of these vectors, thus, knowledge of the scalar value of
the polarizability is sufficient. The power extracted by a dipole
inclusion (dipole moment $p$) from a given incident field $E_{\rm
inc}$ is given by the classical formula \e P_{\rm ext}={1\over
2}{\rm Re}\int_V \_J^*\cdot \_E_{\rm inc}\, dV={1\over 2}{\rm
Re}(-j\omega p^*E_{\rm inc})=-{\omega\over 2}{\rm Im}(\alpha)|E_{\rm
inc}|^2 =-\eta \omega{\rm Im}(\alpha)P_{\rm inc} \l{si}\f (the
harmonic time dependence is in the form $\exp(j\omega t)$). Here
$\_J$ is the volumetric electric current density inside the
particle. This general formula is valid for any dispersive and lossy
particle, assuming only that it is a small dipole particle and
$E_{\rm inc}$ is uniform over its volume. Here and below, the result
in terms of the incident power density $P_{\rm inc}$ holds if the
particle is excited by a propagating plane wave. The corresponding
value of the incident time-averaged power flow density is \e P_{\rm
inc}={1\over 2\eta}|E_{\rm inc}|^2\f where
$\eta=\sqrt{\mu_0/\epsilon_0}$ is the free-space wave impedance. The
power which is scattered (re-radiated) by the particle is given as
the power radiated by the electric dipole $p$:
\begin{equation}\l{scat}
P_{\rm sc}={\mu_0\omega^4|\alpha|^2\over 12 \pi c}|E_{\rm inc}|^2= {k^4\over 6 \pi \epsilon_0^2}|\alpha|^2 P_{\rm inc}
\end{equation}

We can find the general expression for the scattering loss factor in
the dipole polarizability equating the extracted and scattered
powers for the case of a lossless particle (no absorption). This
allows us to find an expression for the imaginary part of the
inverse polarizability (e.g. \cite[eq.~(4.82)]{modeboo}) \e {1\over
\alpha}={{\rm Re}(\alpha)-j{\rm Im}(\alpha)\over |\alpha|^2}\f as \e
{\rm Im}\left({1\over \alpha}\right)=-{{\rm Im}(\alpha)\over
|\alpha|^2}={\mu_0\omega^3\over 6\pi c}={k^3\over 6\pi\epsilon_0}\f
($k$ is the free-space wave number).
This is a classical result which dates back to the work of M. Plank
(1902), see references in \cite{Belov2003}.

For the most general linear particle we can write for the inverse
polarizability
\begin{equation}\l{L}
{1\over \alpha}=\xi'+j\xi'' +j{k^3\over 6\pi\epsilon_0}
\end{equation}
Here the last term is due to the scattering  loss, and the complex
value of $\xi=\xi'+j\xi''$ depends on the particle size, shape,
material, and the frequency. For passive particles the absorption
coefficient $\xi''\ge 0$ (this ensures that the absorbed power is
non-negative, see eq.~\r{sigma_abs} below), and for lossless
particles $\xi''=0$. The real part of the inverse polarizability
$\xi'$ can take any real value. $\xi'=0$ corresponds to the resonant
frequencies of the particle, and $|\xi'|\rightarrow \infty$
corresponds to zero polarizability (particle is not excited at all,
or there is no particle). The scattering-loss term remains the same
also for absorbing particles, because this value depends only on the
frequency and not on the particle parameters.

The absorbed power is the difference between the power extracted by
the particle from the incident field \r{si} and the power scattered
by the same particle into the surrounding space \r{scat}:
\begin{equation}
P_{\rm abs}=-{\omega\over 2}{\rm Im}(\alpha)|E_{\rm inc}|^2
-{\mu_0\omega^4|\alpha|^2\over 12 \pi c}|E_{\rm inc}|^2={{\omega\over 2}\xi''\over{\xi'^2+\left(\xi''+{k^3\over 6\pi\epsilon_0}\right)^2}}
|E_{\rm inc}|
^2
={{k\over \epsilon_0}\xi''\over{\xi'^2+\left(\xi''+{k^3\over 6\pi\epsilon_0}\right)^2}}
P_{\rm inc}
\l{spa}
\end{equation}
Here we have substituted the general expression for the
polarizability $\alpha$ from  \r{L}.

Let us list the corresponding effective cross sections of the
particle, defined as the ratios of these powers to the incident
power density. The extinction cross section: \e \sigma_{\rm
ext}={P_{\rm ext}\over P_{\rm inc}}={k\over \epsilon_0}{\xi''+
{k^3\over 6\pi\epsilon_0} \over \xi'^2+ \left(\xi''+{k^3\over
6\pi\epsilon_0}\right)^2}\l{sigma_ext}\f The absorption cross
section: \e \sigma_{\rm abs}={P_{\rm abs}\over P_{\rm inc}}={k\over
\epsilon_0}{\xi''\over{\xi'^2+\left(\xi''+{k^3\over
6\pi\epsilon_0}\right)^2}}\l{sigma_abs}\f and the total scattering
cross section: \e \sigma_{\rm sc}={P_{\rm sc}\over P_{\rm
inc}}={k^4\over 6 \pi\epsilon_0^2 } {1\over
\xi'^2+\left(\xi''^2+{k^3\over
6\pi\epsilon_0}\right)^2}\l{sigma_sc}\f Finally, the radar
(backscattering) cross section, defined as \e
\sigma=\lim_{D\rightarrow\infty} 4\pi D^2 {|E_{\rm sc}|^2\over
|E_{\rm inc}|^2}\f (here $D$ is the distance from the particle to
the observation point along the direction towards the source) equals
\e \sigma={k^4 |\alpha|^2\over 4\pi \epsilon_0^2}={k^4 \over 4\pi
\epsilon_0^2}{1\over \xi'^2+\left(\xi''^2+{k^3\over
6\pi\epsilon_0}\right)^2}\f Note that it differs from the total
scattering cross section \r{sigma_ext} only by a numerical factor.

In many papers and books, scattering from small particles is
considered based on the quasi-static approximation of the particle
polarizability $\alpha$. Under this approximation, when the loss
parameter decreases, the amplitude of the induced dipole moment can
reach arbitrary large values. This is, naturally, because scattering
from the particle is neglected in the quasi-static approximation,
and there is no fundamental lower limit on the degree of losses. In
this approximation there is no difference between extinction and
absorption cross sections, see \r{sigma_ext} and \r{sigma_abs} in
the assumption that $k\rightarrow 0$. It may be confusing that in
some books (e.g. \cite[Sec.~5.2]{Bohren},
\cite[Eq.~(8.7)]{Capolino}), the \emph{extinction} cross section of
small dipolar particles expressed in terms of the imaginary part of
the polarizability as in \r{si} is called the \emph{absorption}
cross section (although they are indeed not distinguishable in the
quasi-static model).

While for determination of the near-field distribution and for
calculations of the real part of the polarizability the accuracy of
the quasi-static approximation is improving with decreasing the
particle size, the above results for the effective cross section
areas show that neglecting the scattering term proportional to $k^3$
leads to dramatically different results for low-loss resonant
particles. Actually, the quasi-static model of the polarizability
leads to qualitatively wrong (non-physical) results for the
absorption cross section of low-loss particles. In the quasi-static
approximation $\sigma_{\rm abs}$ \r{sigma_abs} at resonance (at
$\xi'=0$) diverges as $1/\xi''$ when the loss parameter
$\xi''\rightarrow 0$, while the full-wave result shows
 that absorption tends to zero as $\sigma_{\rm abs}\sim \xi''$ for small $\xi''$. Actually,  the
 quasi-static approximation can be used for small resonant dipole scatterers
 only when the absorption loss dominates over the scattering loss.
 For plasmonic nanoparticles this means the case of small radius of
 the particle, see formula  \r{xi_plasma} below   (for silver spheres the estimation based on the Drude
 model gives $r\ll 10$ nm \cite{andrei}).
For electrically small low-loss dielectric spheres the quasi-static
approximation is accurate when $|\xi'|\ll k^3/(6\pi \epsilon_0)$,
which is equivalent to \e \left({r\over \lambda}\right)^3\ll {3\over
16\pi^3}\left|{\epsilon+2\over \epsilon-1}\right|\f where $\epsilon$
is the relative permittivity of the sphere material.

\section{Extreme values}

The absorbed power \r{spa} and the absorption cross section
\r{sigma_abs} reach their maxima at \e \xi'=0,\qquad \xi''={k^3\over
6\pi\epsilon_0} \l{condi}\f The maximal possible value of absorbed
power reads, upon substitution of these values,
\begin{equation}\l{mabs}
P_{\rm abs\,\, max}={3\pi\over 2k^2}P_{\rm inc}={3\over 8\pi}\lambda^2P_{\rm inc}
\end{equation}
and the ultimate value of the absorption cross section equals to
${3\over 8\pi}\lambda^2$. When the particle is tuned to absorb the
maximum possible power, the scattered power equals to the absorbed
power. Indeed, substitution of $|\alpha|^2$ with $\xi'=0$ and
$\xi''={k^3\over 6\pi\epsilon_0}$ into the expression for the total
scattered power \r{scat} gives the same result as in \r{mabs}.

Physically, this extremum of absorbed power corresponds to a particle at resonance ($\xi'=0$)
whose loss parameter is such that the amount of absorbed power is
equal to the amount of scattered power. This extremum has the same
meaning as the extremum of power delivered to a load from a
generator when the load impedance is matched to the impedance of the
source. Similarly, in that case half of the power is delivered to
the load and half is dissipated in the internal resistance of the
generator.

We can find the extreme value of the scattering cross section of any
dipole particle from \r{scat} or \r{sigma_sc}. Obviously, scattering
is maximized when  $\xi'=\xi''=0$. As expected, this corresponds to
a lossless particle at its resonance: In this case the induced
dipole moment is maximized, which obviously corresponds to the
maximum scattered power. The corresponding value of the maximum
scattered power reads \e P_{\rm sc\,\, max}={3\over 2\pi}\lambda^2
\, P_{\rm inc}\f and the maximum possible total scattering cross
section equals ${3\over 2\pi } \lambda^2$, four times as large as
the maximum absorption cross section.

The extreme value of the extinction cross section we find maximizing
the extracted power \r{si} or the extinction cross section
\r{sigma_ext}. Also in this case the maximum is reached at
$\xi'=\xi''=0$, the same as for the maximal scattering cross
section. Thus,  the maximum extinction cross section equals to the
maximum scattering cross section ($P_{\rm sc\,\, max}= P_{\rm
ext\,\, max}$), and it is reached when the particle resonates but
does not absorb power.

\subsection{Special cases}

\subsubsection{Small dielectric sphere}

For a special case of a small dielectric sphere (radius $r$) with
the relative permittivity $\epsilon=\epsilon'-j\epsilon''$  the
inverse polarizability reads, with the accuracy up to the
third-order terms  $(kr)^3$ (e.g. \cite[Eq.~(8.10)]{Capolino}) \e
{1\over \alpha}={1\over 4\pi
\epsilon_0r^3}{\epsilon'+2-j\epsilon''-{3\over
5}(\epsilon'-2-j\epsilon'')(kr)^2\over
\epsilon'-1-j\epsilon''}+j{k^3\over 6\pi \epsilon_0}\f Thus,
parameters $\xi'$ and $\xi''$ in \r{L} read \e
\xi'={\left[\epsilon'+2-{3\over 5}
(kr)^2(\epsilon'-2)\right](\epsilon'-1)+\epsilon''^2\left[1-{3\over
5}(kr)^2\right]\over 4\pi\epsilon_0
r^3[(\epsilon'-1)^2+\epsilon''^2]}\f \e  \xi''={3\epsilon''\over
4\pi\epsilon_0 r^3[(\epsilon'-1)^2+\epsilon''^2]}\left(1+{1\over
5}(kr)^2\right)\f

To write the conditions for the maximum absorbed power \r{condi} in
terms of the particle permittivity, we first equate $\xi''$ and
$k^3/(6\pi\epsilon_0)$. In the assumption of small losses, this
gives \e \epsilon''\approx {2\over 9} (kr)^3(\epsilon'-1)^2\f
Substituting into the second condition $\xi'=0$ and keeping the
terms up to the second order, we get \e \epsilon'\approx -2-{3\over
5} (kr)^2\f This finally determines the required loss factor \e
\epsilon''\approx 2(kr)^3\f The conditions for the maximum scattered
power $\xi'=\xi''=0$ read, in the same approximation, \e
\epsilon'\approx -2-{3\over 5} (kr)^2,\qquad \epsilon''=0\f
(lossless plasmonic particle at resonance).

%
%

\subsubsection{Small metal sphere (Drude model)}

For a special case of a metal sphere modelled by the Drude relative
permittivity \e \epsilon=1-{\omega_p^2\over
\omega(\omega-j\Gamma)}\f we have, neglecting the second-order term,

\e \xi'={\omega_p^2-3\omega^2\over
4\pi\epsilon_0r^3\omega_p^2},\qquad \xi''= {3\omega \Gamma\over
4\pi\epsilon_0 r^3 \omega_p^2}\l{xi_plasma}\f Thus, the maximum
absorption corresponds to the frequency equal to the resonant
frequency $\omega=\omega_p/\sqrt{3}$ and to the loss factor equal to
\e \Gamma={2\omega \over 3} (kr)^3\f

\subsubsection{Electrically small dipole antenna}

Let us look at the maximal power received by a short electric dipole
antenna (the arm length $l$). Instead of considering the antenna as
a scatterer in terms of its polarizability, we can find the maximal
received power directly. The electromotive force induced by the
incident field is $E_{\rm inc}l$. To maximize the received power, we
bring the antenna to resonance (by an inductive load) and assume
that the antenna material is perfectly conducting (we look at the
maximum possible received power, so we should maximize the
efficiency to 100\%). Then the input impedance is equal to the
radiation resistance \e R_{\rm rad}={\eta\over
6\pi}(kl)^2\l{R_rad}\f (e.g., \cite{Balanis}).

To maximize the received power, we load  the antenna by a matched
load whose resistance is equal to this radiation resistance. Note
that if we want to maximize the \emph{extracted} power, we do not
need to minimize losses in the antenna itself, as the respective
condition is the equality of the \emph{total} antenna resistance and
the radiation resistance. The current through the load is $I={E_{\rm
inc}l\over 2R_{\rm rad}}$. The power delivered to the load reads \e
P_{\rm abs\,\, max}={1\over 2}|I|^2R_{\rm rad}={6\pi\over 4\eta
k^2}|E_{\rm inc}|^2={3\over 8\pi}\lambda^2 P_{\rm inc}\f \e
\sigma_{\rm abs\,\, max}={3\over 8\pi}\lambda^2\l{Sabs_max}\f is the
maximal value of the effective area of this antenna, the same result
as derived above for arbitrary small particles. For small electric
dipoles this result is known from the antenna theory, e.g.
\cite[eq.~(4-32)]{Balanis}.

The inverse polarizability of a short dipole antenna loaded by a
load impedance $Z_{\rm load}=jX+R$ and tuned close to the resonance
can be written as (e.g., \cite{loaded_dipole}) \e {1\over
\alpha}={j\omega \over l^2}\left(Z_{\rm inp}+Z_{\rm load}\right)\f
where $Z_{\rm inp}$ is the input impedance of the antenna. This
allows us to find \e \xi'=-{\omega\over l^2}\left[{\rm Im}(Z_{\rm
inp})+X\right]\l{xi1_wire} \f \e \xi''={\omega \over l^2}\left[{\rm
Re}(Z_{\rm inp}-R_{\rm rad})+R\right]\l{xi2_wire}\f For a small
antenna, the imaginary part of $Z_{\rm inp}$ is negative (capacitive
reactance) and the real part is the sum of the radiation resistance
\r{R_rad} and the loss resistance in the antenna wires (for
perfectly conducting wires the loss resistance is zero).

\subsection{Generalization using the reciprocity principle}
\label{gain-directivity}

Continuing the example of the dipole antenna, we can write its
maximum effective area (absorption cross section) in terms of the
antenna gain. Because in this case the antenna efficiency is 100\%,
its gain is equal to its directivity, and the directivity of the
small dipole antenna equals to $3/2$ \cite{Balanis}. Thus, we can
re-write relation \r{Sabs_max} as \e \sigma_{\rm abs\,\,
max}={3\over 8\pi}\lambda^2={3\over 2}\left({\lambda^2\over
4\pi}\right)=G\, {\lambda^2\over 4\pi}\f

In the antenna theory it is well known that for any reciprocal
antenna located in a reciprocal environment and matched to its load
the ratio of the antenna gain to its effective area ${G\over
\sigma_{\rm abs}}$ is a universal constant, which is independent
from the antenna type. Thus, the relation \e \sigma_{\rm abs\,\,
max}=G\, {\lambda^2\over 4\pi}\f holds in fact for any reciprocal
antenna matched to its load. This clearly tells the (obvious) fact
that received power can be increased by making the antenna
directive. In terms of the optical language, this tells that the
extinction cross section can be increased over the value given by
\r{Sabs_max}, if higher-order modes are excited in the particle.

In the theory of superdirective antennas it is proven that antenna
directivity can be made arbitrary large if arbitrary antenna current
distributions (with arbitrary fast variations over the antenna
volume) are allowed. This result is usually attributed to C.W. Oseen
(1922), see discussion and references e.g.~in
\cite[Sec.~2.2]{Hansen}. A superdirective antenna is equivalent to
an electrically small particle where many higher-order modes (index
$l\neq 1$ in the Mie theory of scattering by a sphere) are strongly
(resonantly) excited. This requirement obviously presents challenges
in practical realizations of classical superdirective antennas
\cite{Hansen} and recently introduced ``superscatterers'' and
``superabsorbers''
\cite{Lukyanchuk,superscatterer,superabsorber,Fan,superscatterer2}.
Note strong sub-wavelength oscillations of fields clearly visible in
examples presented in \cite{superscatterer}.

\section{Balance between absorption and scattering}

Let us consider the ratio of the absorbed power $P_{\rm abs}$ to the
total extracted power $P_{\rm ext}$. Dividing \r{sigma_abs} by
\r{sigma_ext}, we find \e {P_{\rm abs}\over P_{\rm ext}}
={\xi''\over \xi''+{k^3\over 6\pi\epsilon_0}}\l{PaPt}\f This ratio
can vary from 0 to 1, reaching 1 when $\xi''\rightarrow \infty$, but
of course at this limit there is no absorption nor extinction. Next,
let us find the ratio \e  { P_{\rm abs}\over P_{\rm
sc}}=\xi''{6\pi\epsilon_0\over k^3}\l{PaPs}\f This ratio can take
any value from zero to infinity, but, again, in the limit of
infinity the particle is simply not excited at all, and in the limit
of zero there is no absorption.

\begin{figure}[h]
\centering
\epsfig{file=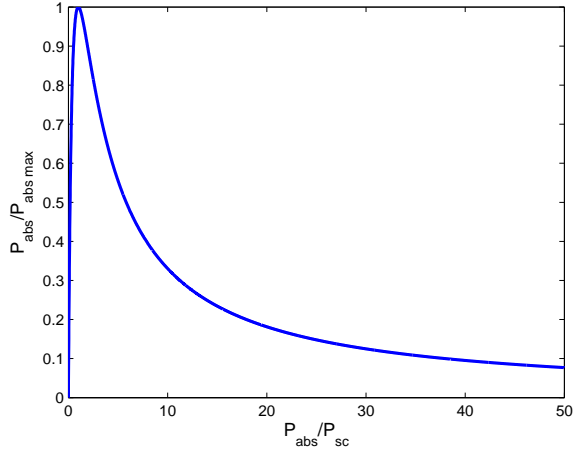, width=0.5\textwidth}
\caption{Dependence of the absorbed power on the ratio of absorbed and scattered powers.
The maximum of absorbed power \r{mabs} is reached at $P_{\rm abs}=P_{\rm sc}$.}
\label{pa_psc}
\end{figure}

For a specific (desired) value of the ratio $P_{\rm abs}/P_{\rm sc}$
at a given frequency there is a corresponding value of the loss
factor $\xi''$ defined by \r{PaPs}. For this value of the power
ratio and the loss factor the absorbed power \r{spa} is defined by
the value of the real part of the inverse polarizability $\xi'$.
Parameter $\xi'^2$ can take any real value from zero (particle at
resonance) to infinity (the polarizability equals zero). The first
case corresponds to the maximum allowed received power for a given
value of the power ratio \r{PaPs}, and the received power is zero in
the last case. The maximal possible received power for a specific
value of the ratio $P_{\rm abs}/P_{\rm sc}$ can be written in the
form \e P_{\rm abs}|_{\xi'=0} = {{ P_{\rm abs}\over P_{\rm sc}}\over
\left(1+ { P_{\rm abs}\over P_{\rm sc}}\right)^2}{3\over
2\pi}\lambda^2 \, P_{\rm inc}\l{lim_ratio} \f Here we have used
\r{PaPs} to express $\xi''$ in terms of the power ratio and the wave
number, and substituted that into \r{spa}. This dependence is
illustrated by Fig.~\ref{pa_psc}.

As was discussed above, the point of the maximal received power
corresponds to $P_{\rm abs}/P_{\rm sc}=1$. To reach the maximum of
the absorbed power for any desired ratio $P_{\rm abs}/P_{\rm sc}$
one should bring the particle to resonance and tune the loss factor
using equation \r{PaPs}. For example, a short dipole antenna can be
tuned to receive maximum possible power for any desired value of
$P_{\rm abs}/P_{\rm sc}$ by loading it with complex impedance
$Z_{\rm load}=j\omega L +R$, where the inductance $L$ is chosen so
that the inductive load compensates the input capacitance of the
antenna ($\xi'$ in \r{xi1_wire} is zero), and the resistor $R$ is
either larger than the radiation resistance \r{R_rad} or smaller
than that. In the first case we can reach any point on the curve to
the right from $P_{\rm abs}/P_{\rm sc}=1$, which corresponds to the
regime of a ``cloaked sensor'' \cite{alu,cloaked_sensor}, where the
ratio $P_{\rm abs}/P_{\rm sc}$ is maximized. Using \r{xi2_wire}, we
can find an elegant design formula for cloaked sensors realized as
simple loaded dipole antennas. Substituting $\xi''$ of a loaded
dipole antenna into \r{PaPs}, we find the required load resistance
to realize a sensor which receives the maximum possible power at a
given level of $P_{\rm abs}/P_{\rm sc}$: \e R= R_{\rm rad} {P_{\rm
abs}\over P_{\rm sc}}\f (we have assumed 100\% efficiency, which
means that ${\rm Re}(Z_{\rm inp}-R_{\rm rad})=0$ in \r{xi2_wire}).
The curve in Fig.~\ref{pa_psc} also shows how much the absorbed
power will drop at any level of achieved scattering reduction. For
example, with increasing the ratio $P_{\rm abs}/P_{\rm sc}$ 40
times,  the absorbed power will drop at least about 10 times. In the
case when $R<R_{\rm rad}$, we reach the regime of the maximized
scattered power for any given level of loss in the antenna.
Similarly, for a plasmonic sphere we can substitute $\xi''$ from
\r{xi_plasma} into \r{PaPs} and find that in order to ensure the
desired value of the ratio between absorbed and scattered powers at
frequency $\omega$, the sphere parameters should satisfy \e {\omega
\Gamma\over \omega_p^2}={2\over 9} (kr)^3\, {P_{\rm abs}\over P_{\rm
sc}}\f

\begin{figure}[h]
\centering
\epsfig{file=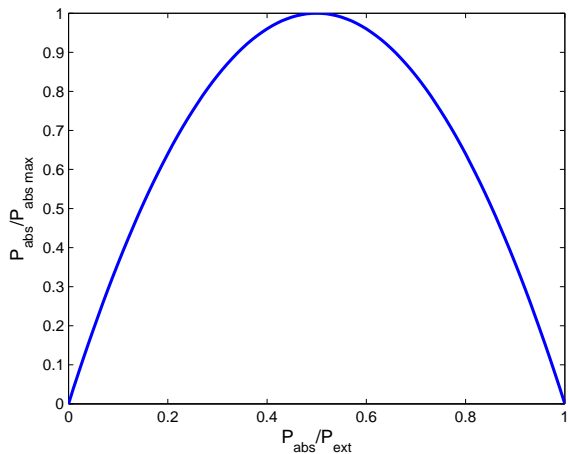, width=0.5\textwidth}
\caption{Dependence of the absorbed power on the ratio of absorbed and extracted powers.
The maximum of absorbed power \r{mabs} is reached at $P_{\rm abs}=P_{\rm ext}/2$.}
\label{pa_pext}
\end{figure}

Using \r{scat} we can find the maximum of the scattered power
reachable for a given level of the ratio \r{PaPs}: \e P_{\rm
sc}|_{\xi'=0} = {1\over 1+{ P_{\rm abs}\over P_{\rm sc}}}{3\over
2\pi}\lambda^2 \, P_{\rm inc}\f This simply tells that scattering is
always reduced with increasing absorption in the particle. In a
similar way we can find the maximal possible absorbed power for a
specific (desired) value of the ratio \r{PaPt}: \e P_{\rm
abs}|_{\xi'=0} = { P_{\rm abs}\over P_{\rm ext}}\left(1- { P_{\rm
abs}\over P_{\rm ext}}\right){3\over 2\pi}\lambda^2 \, P_{\rm inc}
\l{lim_ratio1} \f This dependence is shown in Fig.~\ref{pa_pext}. As
discussed above, the ultimate maximum of the received power \r{mabs}
corresponds to $P_{\rm abs}=P_{\rm sc}$ in \r{lim_ratio} and
\r{lim_ratio1}.

More careful and detailed analysis of these limits and their
implications for the design of small weakly-scattering antennas
(\emph{cloaked sensors}) can be found in paper \cite{alu}.

Considering the special case of a plasmonic nanoparticle, one can
substitute $\xi''$ for a plasmonic nanosphere from \r{xi_plasma} and
study how these ratios depend on the sphere radius and metal
parameters. Basically, $\xi''$ is inversely proportional to $r^3$,
thus, for small spheres it becomes large, both ratios grow, and it
looks like the spheres absorb more. But of course the absorbed power
in fact becomes small for large $\xi''$, as we know from \r{spa}.

\section{Maximal power absorbed and scattered by small particles in planar regular arrays}

Let us consider an infinite two-dimensional periodical array (square
unit cells) of arbitrary electrically small particles and study the
same limits for particles in regular arrays. This issue was
considered in \cite{Pozar2009} in terms of the antenna array theory.
We assume that the array is excited by a normally incident plane
wave and the array period $a$ is smaller than $\lambda$, so that no
diffraction lobes appear. Similarly to the previous sections, we
assume that the particles are small dipole particles excited by
electric fields at the position of the particles. The periodical
grid can be conveniently modelled by its grid impedance $Z_g$
\cite[Ch.~4]{modeboo}, which relates the total (surface-averaged)
electric field $E_{\rm tot}$ in the array plane and the averaged
surface current density $J={j\omega p\over a^2}$: \e E_{\rm tot}=Z_g
J\l{Z_gJ}\f In terms of the polarizability of the individual
particles in free space $\alpha$, defined in \r{L}, the grid
impedance reads \cite[Eq.~(4.95)]{modeboo} \e
Z_g=Z_g'+jZ_g''=-j{a^2\over
\omega}\left(\hat\xi'+j\xi''\right)\l{Zg}\f where $a$ is the array
period and the hat indicates that the real part of the inverse
polarizability is modified due to reactive-field interactions
between the particles in the infinite array. For electrically dense
arrays of small particles this interaction can be estimated
analytically as \e \hat\xi'=\xi'-{\rm Re}(\beta)\approx
\xi'-{0.36\over \epsilon_0 a^3}\f (for our consideration here the
value of the real part of the interaction constant $\beta$ will be
not important).

Next we write the reflection and transmission coefficients (e.g.,
\cite[Eq.~(4.43)]{modeboo} \e R=-{{\eta\over 2}\over {\eta\over
2}+Z_g}\l{R_array}\f \e T={Z_g\over {\eta\over 2}+Z_g}\l{T_array} \f
where, as before, $\eta=\sqrt{\mu_0\over \epsilon_0}$. The
absorption coefficient reads \e L=1-|R|^2-|T|^2={\eta Z_g'\over
Z_g''^2+\left({\eta\over 2}+Z_g'\right)^2}\l{LL}\f and the power
absorbed by one unit area of the array is \e P_{\rm abs}=LP_{\rm
inc}={\eta Z_g'\over Z_g''^2+\left({\eta\over
2}+Z_g'\right)^2}P_{\rm inc}={{k\over \epsilon_0 a^2}\, \xi''\over
{\hat\xi}'^2+\left({k\over 2\epsilon_0 a^2}+\xi''\right)^2}P_{\rm
inc} \l{LLa}\f

Obviously, the highest value of the absorbed power is $P_{\rm abs\
max}={1\over 2}P_{\rm inc}$, and it is reached if \e Z_g''=0,\qquad
Z_g'={\eta\over 2}\f or, in terms of the particle polarizability, \e
\hat\xi'=0,\qquad \xi''={k \over 2\epsilon_0 a^2}\l{optarray}\f The
first condition physically means that the particles are at resonance
(including the reactive coupling with all the other particles in the
array). The second condition means that the absorption in the array
is optimized so that the absorbed power is equal to the re-radiated
power (this value of $\xi''$ is equal to the term in the imaginary
part of the interaction constant which measures the plane-wave power
radiated  by the unit area of the infinite array, see
\cite[Eq.~(4.88)]{modeboo}).

Let us discuss the balance between the absorbed and re-radiated
powers in more detail. We can find the power scattered by a unit
area of the infinite array in the same way as for a single particle:
as the difference of the power extracted from the incident field and
the absorbed power. The extracted power is, similarly to \r{si}, \e
P_{\rm ext}={1\over 2}{\rm Re}\,(J^*E_{\rm inc})\l{Pext_array}\f
where $J$ is the surface current density. Substituting the total
tangential surface-averaged electric field in the array plane in
terms of the incident field and the plane-wave field created by the
induced surface currents as $E_{\rm tot}=E_{\rm inc}-{\eta\over 2}J$
into \r{Z_gJ}, we can express the induced current density in terms
of the incident field: \e J={1\over {Z_g+{\eta\over 2}}}E_{\rm
inc}\f Substitution into \r{Pext_array} gives the extracted power \e
P_{\rm ext}= \eta{Z_g'+{\eta\over 2} \over Z_g''^2+\left({\eta\over
2}+Z_g'\right)^2}P_{\rm inc} ={k\over \epsilon_0 a^2}{\xi''+{k\over
2\epsilon_0 a^2} \over {\hat\xi}'^2+\left(\xi''+ {k\over 2\epsilon_0
a^2}\right)^2}P_{\rm inc} \f and the scattered (re-radiated) power
\e P_{\rm sc} =P_{\rm ext}-P_{\rm abs}={{\eta^2\over 2} \over
Z_g''^2+\left({\eta\over 2}+Z_g'\right)^2}P_{\rm inc}= {1\over
2}\left({k\over \epsilon_0 a^2}\right)^2{1 \over
{\hat\xi}'^2+\left(\xi''+ {k\over 2\epsilon_0 a^2}\right)^2}P_{\rm
inc} \l{Psc_array}\f (we have used formula \r{LLa} for the absorbed
power). Note that this definition implies that the array re-radiates
symmetrically in the backward and forward directions (the same
result \r{Psc_array} for the scattered power $P_{\rm sc}$ can be
obtained as $P_{\rm sc}=2|R|^2 P_{\rm inc}$, see \r{R_array}). The
power which is actually propagating in the space behind the array is
the result of interference of this re-radiated plane wave and the
incident plane wave, and it is given by $|T|^2 P_{\rm inc}$. For
this reason, the re-radiated power can take values between zero and
$2P_{\rm inc}$. The last value corresponds to the case of total
reflection, where the array generates reflected power which is equal
to the incident power \emph{and} creates secondary field behind the
array which cancels the incident field there. This is the same
definition as the definition of the total scattering cross section
in the theory of diffraction (the total scattering cross section of
a large conducting body is equal to the double geometrical cross
section).

Now we can study the ratios \e {P_{\rm abs}\over P_{\rm abs}+P_{\rm
sc}}={P_{\rm abs}\over P_{\rm ext}}={Z_g'\over Z_g' +{\eta\over 2}}={\xi''\over \xi''+{ k \over
2\epsilon_0 a^2}}\f and \e {P_{\rm abs}\over P_{\rm sc}}=
{2Z_g'\over \eta}=\xi''{2\epsilon_0 a^2\over k}\l{PaPs_array}\f (we
have used the relation between the surface resistance of the
homogenized sheet and the loss factor of a single particle
$Z_g'=\xi''a^2/ \omega$ \r{Zg}, \cite[eq.~(4.95)]{modeboo}).

These results are analogous to those for single particles in free
space \r{PaPt} and \r{PaPs}: The single-particle radiation damping
factor $k^3\over 6\pi \epsilon_0$ is replaced by the term ${k \over
2\epsilon_0 a^2}$ which measures the plane-wave power radiated from
a unit area of the infinite array. These are the two terms of the
imaginary part of the interaction constant for infinite arrays of
dipole particles, see \cite[Eq.~(4.89)]{modeboo}. We can make a
similar conclusion about the ratio between the absorbed and
scattered powers: Because the surface resistance $Z_g'$ and take any
value between zero (lossless particles) and infinity (no particles
at all), the ratio \r{PaPs_array} can take any value between zero
and infinity. On the other hand, the value of the absorbed power
\r{LLa} cannot be larger than one half of the incident power, and it
tends to zero when ratio \r{PaPs_array} tends to infinity.

In the same way as for a single particle we can find the maximum
received power per unit area achievable at a given value of the
ratio \r{PaPs_array}. Using \r{LLa} together with \r{PaPs_array}, we
arrive to \e P_{\rm abs}|_{Z_g''=0}= {{P_{\rm abs}\over P_{\rm
sc}}\over \left(1+ {P_{\rm abs}\over P_{\rm sc}}\right)^2}\, 2
P_{\rm inc}\f Tuning the particle reactance, the absorbed power
changes between zero and the extremal value given by the above
relation. We see again that the ultimate maximum is $0.5\,P_{\rm
inc}$, reachable for $P_{\rm abs}=P_{\rm sc}$. One can also study
how the scattering efficiency is limited by losses in the particles,
calculating the maximal scattered power for a fixed level of the
ratio \r{PaPs_array}: \e P_{\rm sc}|_{Z_g''=0}= {1\over \left(1+
{P_{\rm abs}\over P_{\rm sc}}\right)^2}\, 2 P_{\rm inc}\f

Let us look at how much power is absorbed by each particle in the
array and compare  with the absorption by the same particle in free
space \r{spa}. The power absorbed by one unit area (1 m$^2$) of the
array we get multiplying $L$ in \r{LL} by the incident power density
$P_{\rm inc}={1\over 2\eta }|E_{\rm inc}|^2$. Next, the power per
particle we get multiplying by the unit-cell area $a^2$. The result
is \e P_{\rm abs}|_{\rm per \ particle}={{a^2 \over 2}Z_g'\over
Z_g''^2+\left({\eta\over 2}+Z_g'\right)^2}|E_{\rm inc}|^2
={{\omega\over 2}\xi''\over{\hat\xi'^2+\left(\xi''+{\eta\omega\over
2 a^2}\right)^2}}|E_{\rm inc}|^2 ={{\omega\over
2}\xi''\over{\hat\xi'^2+\left(\xi''+{k\over 2
a^2\epsilon_0}\right)^2}}|E_{\rm inc}|^2\f (we have substituted the
value of $Z_g$ in terms of the particle polarizability using
\r{Zg}). Substitution of the optimal value of the particle
polarizability \r{optarray} gives the maximum power received by one
particle \e P_{\rm abs\,\, max}={a^2\over 4\eta}|E_{\rm
inc}|^2={a^2\over 2}P_{\rm inc}\f Thus, the extreme value of the
effective area of each particle acting as a receiving antenna in an
infinite array is \e \sigma_{\rm abs\,\, max}={a^2\over 2}\f This
agrees with the fact that an infinite array of electric dipole
antennas can receive at maximum one half of the incident power (the
area of the unit cell is $a^2$, but the effective receiving area of
each dipole is $a^2/2$). It is interesting to observe that particles
in the grid absorb as much as the same  particles in free space when
\e ka=\sqrt{3\pi}\l{same} \f For \emph{denser} grids each particle
dissipates smaller power than the same particle in free space
illuminated by the same incident field.

The power scattered by each particle in an infinite grid we can find
multiplying \r{Psc_array} by the unit-cell area $a^2$: \e P_{\rm
sc}|_{\rm per \ particle}={{a^2\eta^2\over 2} \over
Z_g''^2+\left({\eta\over 2}+Z_g'\right)^2}P_{\rm inc}\f The maximal
scattered power is achieved when $Z_g=0$, and it is equal to $2a^2
P_{\rm inc}$: The effective scattering area of each unit cell is
\emph{twice as large} as the physical area. This means that the
array fully reflects the incident plane wave \emph{and} creates a
complete shadow behind the array (the reflection coefficient $R=-1$
and the transmission coefficient $T=0$, see \r{R_array} and
\r{T_array}). In terms of the particle polarizability this regime
corresponds to lossless ($\xi''=0$) and resonant (in the array)
particles ($\hat\xi'=0$). Continuing the comparison of properties of
individual particles in free space and the same particles in regular
arrays, we see that also the scattered powers are the same if
\r{same} is satisfied. Note that the ratio between the maximum
scattered and received powers for regular grids is the same as for
individual particles in free space (four).

The largest distance between small particles in a regular array such
that no grating lobes still appear equals $\lambda$ (for the normal
incidence). At this period, the extreme values of absorption and
scattering cross sections per particle in the array become
$\sigma_{\rm abs \ max}={1\over 2}\lambda^2$, $\sigma_{\rm abs \
max}=2\lambda^2$, which are larger values than for the same
particles in free space. Thus, coherent interactions in the grid can
enhance interactions of the particles with the fields. On the other
hand, for dense arrays ($a\ll \lambda$), each particle in the array
absorbs and scatters less than it would do individually in free
space.

Finally we note that adding an array of similarly optimized magnetic
dipoles to the array we can receive 100\% of the incident power,
which is also a known result \cite{Pozar_array,Pozar2009}.

\section{Conclusion}

The limitations on the extinction, absorption, and scattering
properties of arbitrary small scatterers or antennas come from the
fact that the maximum amplitude of the induced dipole moment is
limited by the inevitable scattering loss. Since the scattering loss
factor is the same for all dipole scatterers (it depends only on the
frequency), the limitations are very general. The only assumptions
(besides linearity) are that the particle is small (reacts as an
electric dipole) and that there is no bi-anisotropy and no magnetic
polarization. Absorption and scattering by a magnetic dipole
particle obeys the same rules, as follows from duality (the
corresponding formulas can be obtained replacing $\epsilon_0$ by
$\mu_0$ and \emph{vice versa}. Combining a magnetic dipole with an
electric dipole, we can double the received power. More general
bi-anisotropic particles are considered in recent paper
\cite{younes}.

\section*{Acknowledgements}

Most of this  text was written when the author was visiting the
Technical University of Denmark (DTU Fotonik), in January--April
2013. Financial support from the Otto M{\o}nsted Foundation and
Aalto School of Electrical Engineering is appreciated. Special
thanks to the host, prof. A. Lavrinenko, to Dr. A. Andryieuski for
sharing his knowledge of the optical literature and to prof. O.
Breinbjerg for explaining that the reciprocity relation between the
antenna gain and effective area can be used to find the maximum
received power (Section~\ref{gain-directivity}). The last stages of
this work got inspirations from discussions with prof. A. Al\`u
(University of Texas at Austin), who kindly shared preprint
\cite{alu} with the author. Useful discussions with prof. C.
Simovski (Aalto University) of extinction by particles in regular
arrays are also acknowledged.

\end{document}